\documentclass[twocolumn,numbers]{elsarticle}

\usepackage[version=3]{mhchem}

\usepackage{balance}
\usepackage{mathptmx}
\usepackage{sectsty}
\usepackage[export]{adjustbox}
\usepackage{lmodern}
\usepackage{graphicx} 
\usepackage{subfig}
\newcommand{\keywords}[1]{{#1}}

\usepackage{lastpage}
\usepackage[format=plain,justification=justified,singlelinecheck=false,font={stretch=1.125,small,sf},labelfont=bf,labelsep=space]{caption}
\usepackage{float}
\usepackage{fancyhdr}
\usepackage{fnpos}
\usepackage[english]{babel}

\usepackage{array}
\usepackage{droidsans}
\usepackage{charter}
\usepackage[T1]{fontenc}

\usepackage{setspace}
\usepackage[compact]{titlesec}

\usepackage[breaklinks=true]{hyperref}

\usepackage{mhchem}
\usepackage{lipsum}
\usepackage{tgtermes}
\usepackage{tikz}
\usepackage[]{mhchem}
\usepackage[letterpaper, portrait, margin=0.60in]{geometry}
\usepackage{hhline}
\usepackage{float}
\usepackage{multirow}
\usepackage{graphicx,booktabs}
\usepackage{tabularx}
\usepackage{xfrac}
\usepackage[compact]{titlesec}
\usepackage{wrapfig,booktabs}
\usepackage{mwe}
\usepackage{adjustbox}
\usepackage{caption}
\usepackage{makecell}
\usepackage{rotating}
\usepackage{overpic}
\usepackage{pict2e}
\usepackage{amsmath,array, amssymb}
\usepackage{xcolor}
\usepackage{tcolorbox}
\usepackage{calrsfs}
\usepackage{longtable}
\usepackage{ragged2e}
\usepackage{siunitx}
\usepackage{tikz}
\usepackage{svg}
\usepackage{booktabs}

\definecolor{cream}{RGB}{222,217,201}

\begin{document}

\pagestyle{fancy}
\thispagestyle{plain}
\fancypagestyle{plain}{
\renewcommand{\headrulewidth}{0pt}
}

\makeFNbottom
\makeatletter
\renewcommand\LARGE{\@setfontsize\LARGE{15pt}{17}}
\renewcommand\Large{\@setfontsize\Large{12pt}{14}}
\renewcommand\large{\@setfontsize\large{10pt}{12}}
\renewcommand\footnotesize{\@setfontsize\footnotesize{7pt}{10}}
\makeatother

\renewcommand{\thefootnote}{\fnsymbol{footnote}}
\renewcommand\footnoterule{\vspace*{1pt}%
\color{cream}\hrule width 3.5in height 0.4pt \color{black}\vspace*{5pt}} 
\setcounter{secnumdepth}{5}

\makeatletter 
\renewcommand\@biblabel[1]{#1}            
\renewcommand\@makefntext[1]%
{\noindent\makebox[0pt][r]{\@thefnmark\,}#1}
\makeatother 
\renewcommand{\figurename}{\small{Fig.}~}
\sectionfont{\sffamily\Large}
\subsectionfont{\normalsize}
\subsubsectionfont{\bf}
\setstretch{1.125} 
\setlength{\skip\footins}{0.8cm}
\setlength{\footnotesep}{0.25cm}
\setlength{\jot}{10pt}
\titlespacing*{\section}{0pt}{4pt}{4pt}
\titlespacing*{\subsection}{0pt}{15pt}{1pt}


\fancyfoot{}
\fancyfoot[LO,R]{\vspace{-7.1pt}\includegraphics[height=9pt]{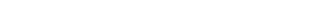}}
\fancyfoot[CO]{\vspace{-7.1pt}\hspace{13.2cm}\includegraphics{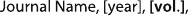}}
\fancyfoot[C]{\vspace{-7.2pt}\hspace{-14.2cm}\includegraphics{head_foot/RF}}
\fancyfoot[RO]{\footnotesize{\sffamily{1--\pageref{LastPage} ~\textbar  \hspace{2pt}\thepage}}}
\fancyfoot[L]{\footnotesize{\sffamily{\thepage~\textbar\hspace{3.45cm} 1--\pageref{LastPage}}}}
\fancyhead{}
\renewcommand{\headrulewidth}{0pt} 
\renewcommand{\footrulewidth}{0pt}
\setlength{\arrayrulewidth}{1pt}
\setlength{\columnsep}{6.5mm}
\setlength\bibsep{1pt}

\makeatletter 
\newlength{\figrulesep} 
\setlength{\figrulesep}{0.5\textfloatsep} 

\newcommand{\topfigrule}{\vspace*{-1pt}%
\noindent{\color{cream}\rule[-\figrulesep]{\columnwidth}{1.5pt}} }

\newcommand{\botfigrule}{\vspace*{-2pt}%
\noindent{\color{cream}\rule[\figrulesep]{\columnwidth}{1.5pt}} }

\newcommand{\dblfigrule}{\vspace*{-1pt}%
\noindent{\color{cream}\rule[-\figrulesep]{\textwidth}{1.5pt}} }

\makeatother

\twocolumn[
  \begin{@twocolumnfalse}
{{\includegraphics[height=55pt]{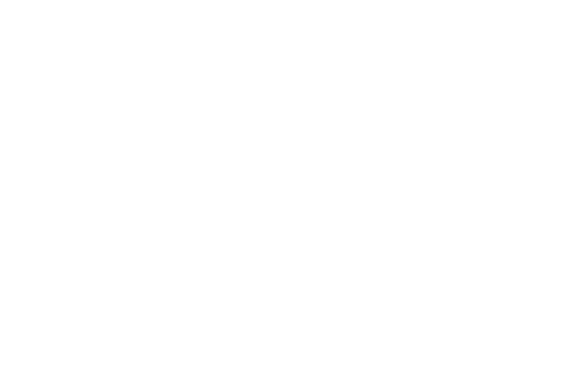}}\\[1ex]
\includegraphics[width=18.5cm]{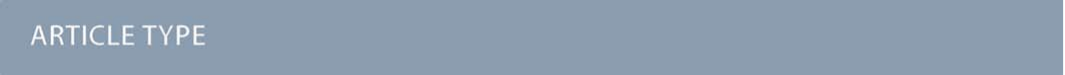}}\par
\vspace{1em}
\sffamily
\begin{tabular}{m{4.5cm} p{13.5cm} }

\includegraphics{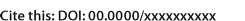} & \noindent\LARGE{\textbf{Data-Driven Insights into Composition–Property Relationships in FCC High Entropy Alloys}} \\
\vspace{0.3cm} & \vspace{0.3cm} \\

 & \noindent\large{Nicolás Flores,\textit{$^{a\dag}$} Daniel Salas Mula,\textit{$^{a}$} Wenle Xu,\textit{$^{a}$}} Sahu Bibhu,\textit{$^{a}$} Daniel Lewis,\textit{$^{a}$} Alexandra Eve Salinas,\textit{$^{b}$} Samantha Mitra,\textit{$^{c}$} Raj Mahat,\textit{$^{c}$} Surya R. Kalidindi,\textit{$^{c}$} Justin Wilkerson,\textit{$^{b}$} James Paramore,\textit{$^{a,d,e}$} Ankit Srivastiva,\textit{$^{a}$} George Pharr,\textit{$^{a}$} Douglas Allaire,\textit{$^{b}$} Ibrahim Karaman,\textit{$^{a}$} Brady Butler,\textit{$^{a,e}$} Vahid Attari,\textit{$^{a}$} and Raymundo Arróyave\textit{$^{a}$}\\

\includegraphics{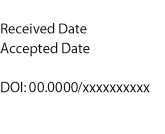} & \noindent\normalsize{Structural High Entropy Alloys (HEAs) are crucial in advancing technology across various sectors, including aerospace, automotive, and defense industries. However, the scarcity of integrated chemistry, process, structure, and property data presents significant challenges for predictive property modeling. Given the vast design space of these alloys, uncovering the underlying patterns is essential yet difficult, requiring advanced methods capable of learning from limited and heterogeneous datasets. This work presents several sensitivity analyses, highlighting key elemental contributions to mechanical behavior, including insights into the compositional factors associated with brittle and fractured responses observed during nanoindentation testing in the BIRDSHOT center Ni-Co-Fe-Cr-V-Mn-Cu-Al system dataset. Several encoder–decoder–based chemistry–property models, carefully tuned through Bayesian multi–objective hyperparameter optimization, are evaluated for mapping alloy composition to six mechanical properties. The models achieve competitive or superior performance to conventional regressors across all properties, particularly for yield strength and the UTS/YS ratio, demonstrating their effectiveness in capturing complex composition–property relationships.
}\\

\end{tabular}

 \end{@twocolumnfalse} \vspace{0.6cm}
]

 \vspace{1.2cm}


\renewcommand*\rmdefault{bch}\normalfont\upshape
\rmfamily
\vspace{-1cm}

{\renewcommand{\thefootnote}{}\footnotemark}
\footnotetext{\textit{$^{a}$~Materials Science and Engineering Department, Texas A\&M University, College Station, TX.}}
\footnotetext{\textit{$^{b}$~Mechanical Engineering Department, Texas A\&M University, College Station, TX.}}
\footnotetext{\textit{$^{c}$~School of Materials Science and Engineering, Georgia 
Institute of Technology, Atlanta, GA.}}
\footnotetext{\textit{$^{d}$~Bush Combat Development Complex, Texas A\&M University System, College Station, TX }}
\footnotetext{\textit{$^{e}$~DEVCOM Army Research Laboratory, ARL South at TAMU, College Station, TX  }}

\footnotetext{\dag~Corresponding Author, Nicolás Flores, nflores6930@tamu.edu}


\sloppy


The discovery and design of high–performance materials for extreme environments, such as those encountered in aerospace, defense, and nuclear applications, requires navigating vast compositional spaces with limited experimental data. High–entropy alloys (HEAs), characterized by their multi–principal element compositions, offer potentially exceptional combinations of strength, thermal stability, and oxidation resistance~\cite{cantor2004microstructural, yeh2004nanostructured}. However, understanding and predicting their behavior remains challenging due to complex, non–linear, and often poorly understood relationships between chemistry, processing, structure, and properties. These challenges are compounded by the scarcity of available datasets, which limit the effectiveness of traditional modeling approaches~\cite{mulukutla2024illustrating,borg2020expanded,allen2024performance}.

\begin{figure*}[t]
    \centering    
    \includegraphics[width=.9\textwidth]{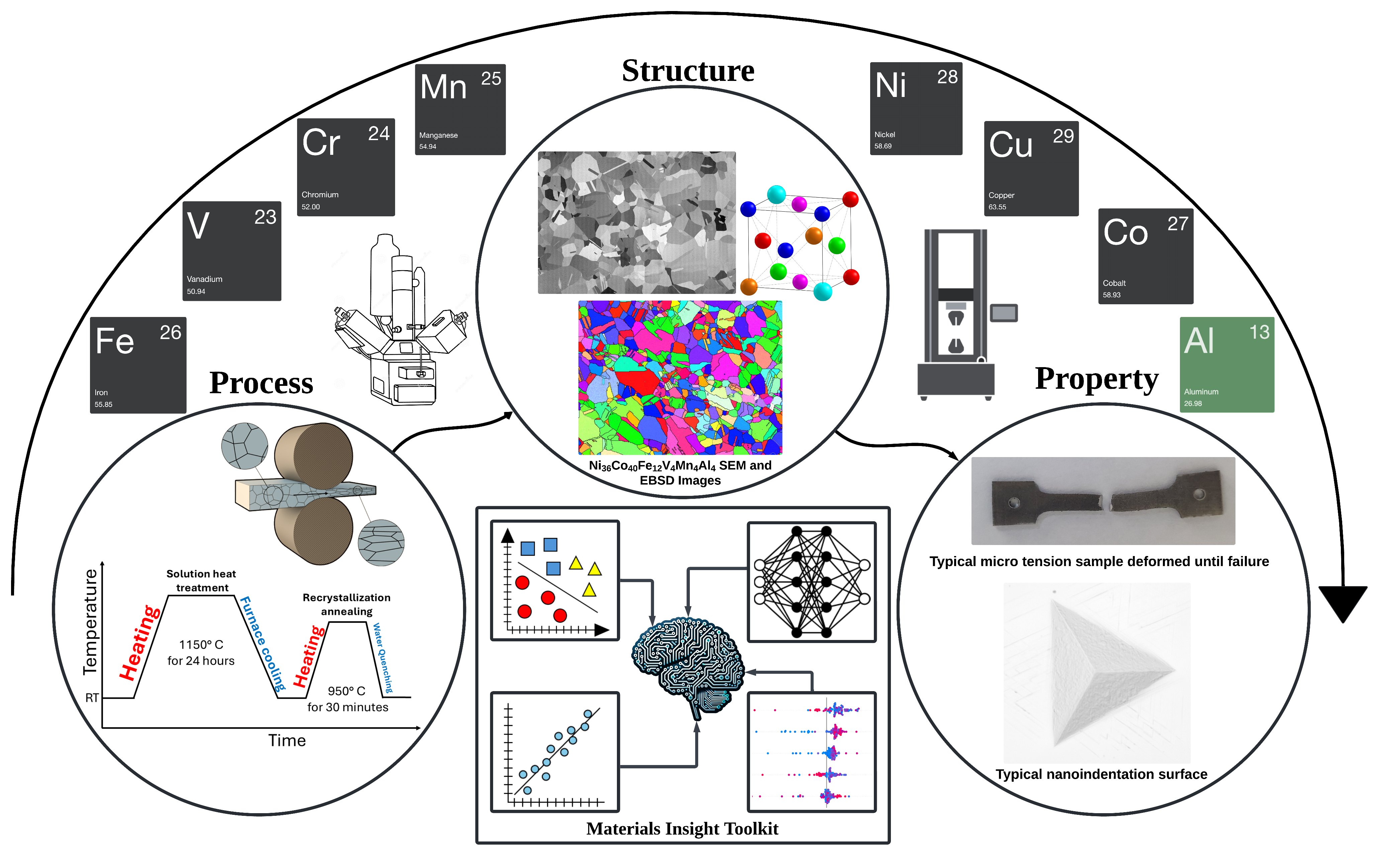}
    \caption{\textbf{Integrated machine learning workflow for high–entropy alloy discovery.} Integrated framework linking alloy composition, processing routes, experimental characterization, and machine learning tools to enable data–driven materials discovery and property optimization.} 
    \label{fig:HEA_ML_Workflow}
\end{figure*}

Machine learning (ML) offers a promising path forward by enabling the extraction of complex, non–linear relationships between alloy composition and mechanical properties from small yet high–dimensional HEA datasets that are often incomplete, inconsistent, or sparsely populated with experimental measurements. Yet, the practical application of ML to HEAs is constrained by the extremely limited volume of available high–quality experimental data relative to the vast and continuous compositional space these alloys span. This data sparsity, combined with the complexity and limited interpretability of many ML models, poses significant challenges for developing reliable and generalizable property predictions. In this work, we develop an encoder–decoder model based on regularized dense networks for learning composition–property relationships in the Ni-Co-Fe-Cr-V-Mn-Cu-Al system. Trained on limited data, the model provides a robust approach for decoding the nonlinearities in the high–dimensional HEA space.

Figure~\ref{fig:HEA_ML_Workflow} illustrates a modern discovery framework that brings together processing conditions, microstructural features (via SEM, EBSD, and XRD), and mechanical property measurements (nanoindentation and tensile testing) into a unified, ML–driven platform. This work focuses on understanding the chemistry–property relations via the Materials Insight Toolkit (bottom–center), which learns composition–property relationships to support alloy design even when data is limited. Within this toolkit, the encoder–decoder model is trained exclusively on compositional inputs to predict key mechanical properties and serves a critical upstream role: providing ML–based prior estimates to guide the primary Bayesian Optimization (BO) framework employed for experimental alloy discovery at the BIRDSHOT center. By learning complex composition–property relationships from limited data, the model enables informed exploration of the design space, helping prioritize regions likely to yield desirable property combinations. This integration of a predictive prior into the BO loop enhances efficiency, reduces the number of costly experiments, and accelerates discovery within the vast compositional landscape of HEAs.

\begin{figure*}[!ht]
    \centering    
    \includegraphics[width=.9\textwidth]{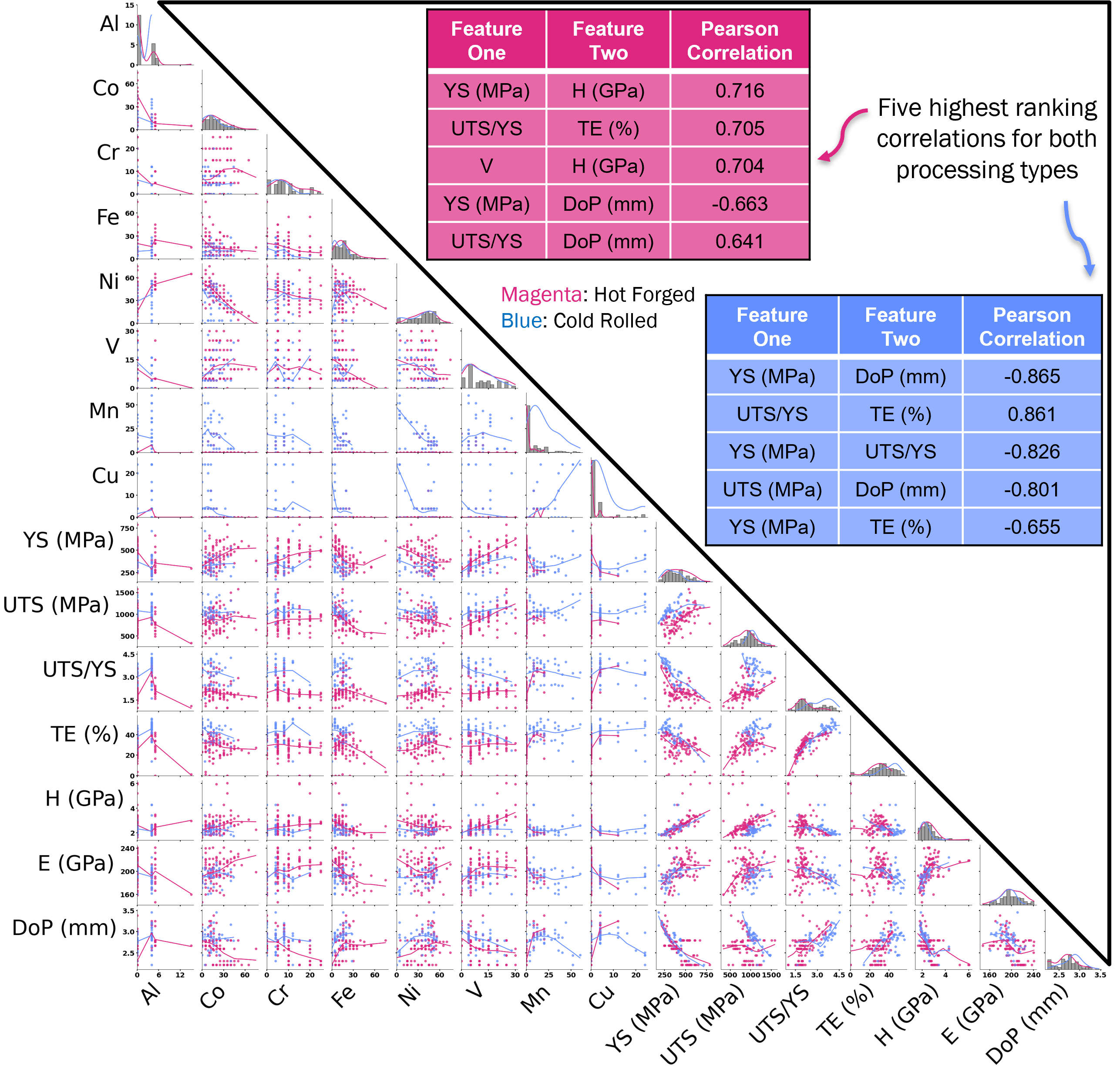}
    
    \caption{\textbf{Comprehensive trend and distribution analysis of compositional and mechanical property features across the Ni-Co-Fe-Cr-V-Mn-Cu-Al high–entropy alloy dataset.} The matrix combines scatter plots (lower triangle), kernel density estimates (diagonal), and Pearson correlation coefficients (upper triangle) to visualize relationships among elements and properties such as yield strength (YS), ultimate tensile strength (UTS), UTS/YS ratio, tension elongation (TE), nanoindentation hardness (H), modulus (E), and simulated depth of penetration (DoP). A realization of each is implemented for each of the two processing types in the data, revealing distinct relationships in each.  Notably, vanadium strongly correlates with increased hardness in hot forged samples, while UTS/YS ratio has a strong inverse relationship with TE.
    }    
    \label{fig:Correlation Matrix}
\end{figure*}

To support the development of this ML–guided discovery framework, a two–year high–throughput experimental campaign has been conducted by the BIRDSHOT center at Texas A\&M University, yielding a foundational dataset for model training and validation. A total of 142 HEAs within the Ni-Co-Fe-Cr-V-Mn-Cu-Al system were designed and synthesized following a Bayesian discovery framework~\cite{khatamsaz2022multi, khatamsaz2023bayesian,hastings2024interoperable}. All compositions were prepared via vacuum arc melting, with year one samples hot forged and year two samples cold rolled and recrystallized. In this work, we focus solely on chemistry, rather than microstructural attributes, for model building. To address missing property values resulting from experimental limitations, a 1-nearest neighbor machine learning model was used to interpolate incomplete entries, ensuring a consistent dataset suitable for learning composition–property relationships under data–scarce conditions \cite{demvsar2004orange}. This interpolation was performed on 11 Yield Strength (YS), Ultimate Tensile Strength (UTS), and Tension Elongation (TE) data, 15 hardness data, and 14 modulus data. 6 data points were removed for tension elongation predictive modeling as these exhibited brittle failure and hindered model tuning. Because depth of penetration (DoP) simulations did not begin until later campaigns, 90 interpolations were performed on this dataset to gain insights into how this property correlates with other features in our dataset. 

\begin{figure*}[!ht]
    \centering

    \makebox[\textwidth][c]{  
        \subfloat[Six nanoindentation samples showing plain(left column), slip traces and pileup (center column), and brittle cracked indentation responses (right column).]{%
            \includegraphics[ width=0.4\textwidth]{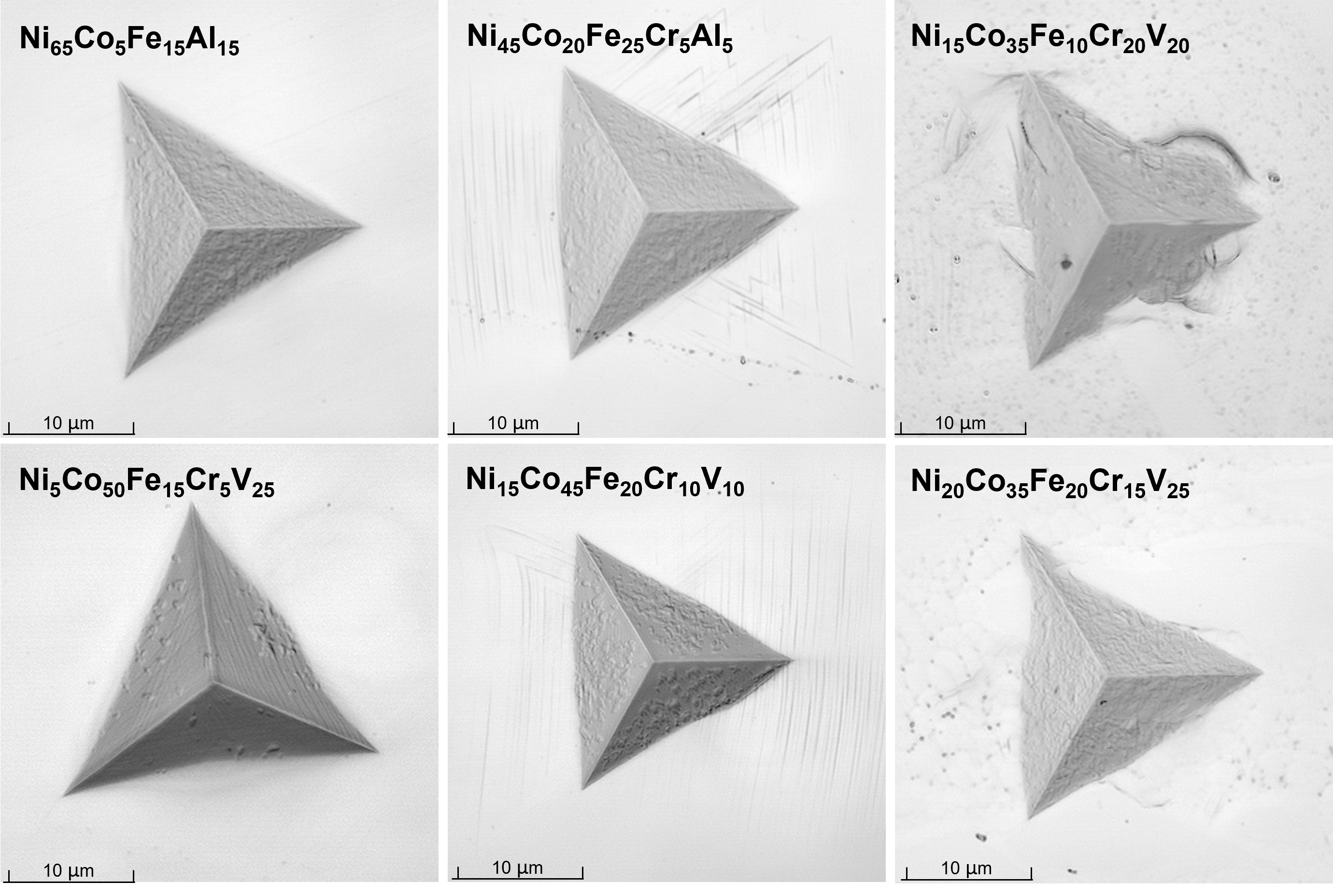}
        }%
        
        \hspace{2em}
        \subfloat[Composition feature importance for influencing brittle nanoindentation results, sorted by importance.]{%
            \includegraphics[width=0.5\textwidth]{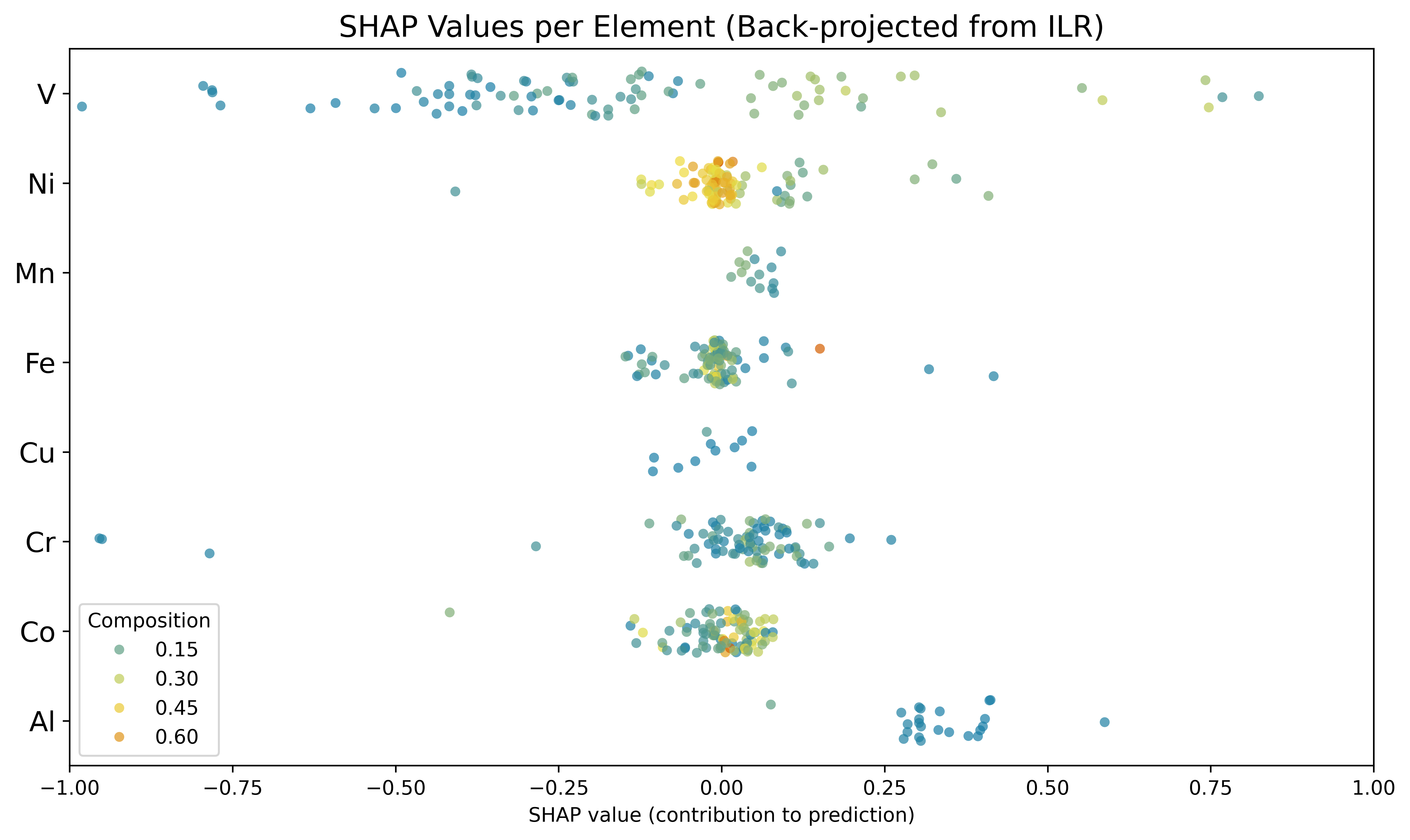}
        }
    }

    \vspace{1ex}  

    \makebox[\textwidth][c]{  
        \subfloat[Principal Component Analysis (PCA) across the composition space.]{%
            \includegraphics[width=0.45\textwidth]{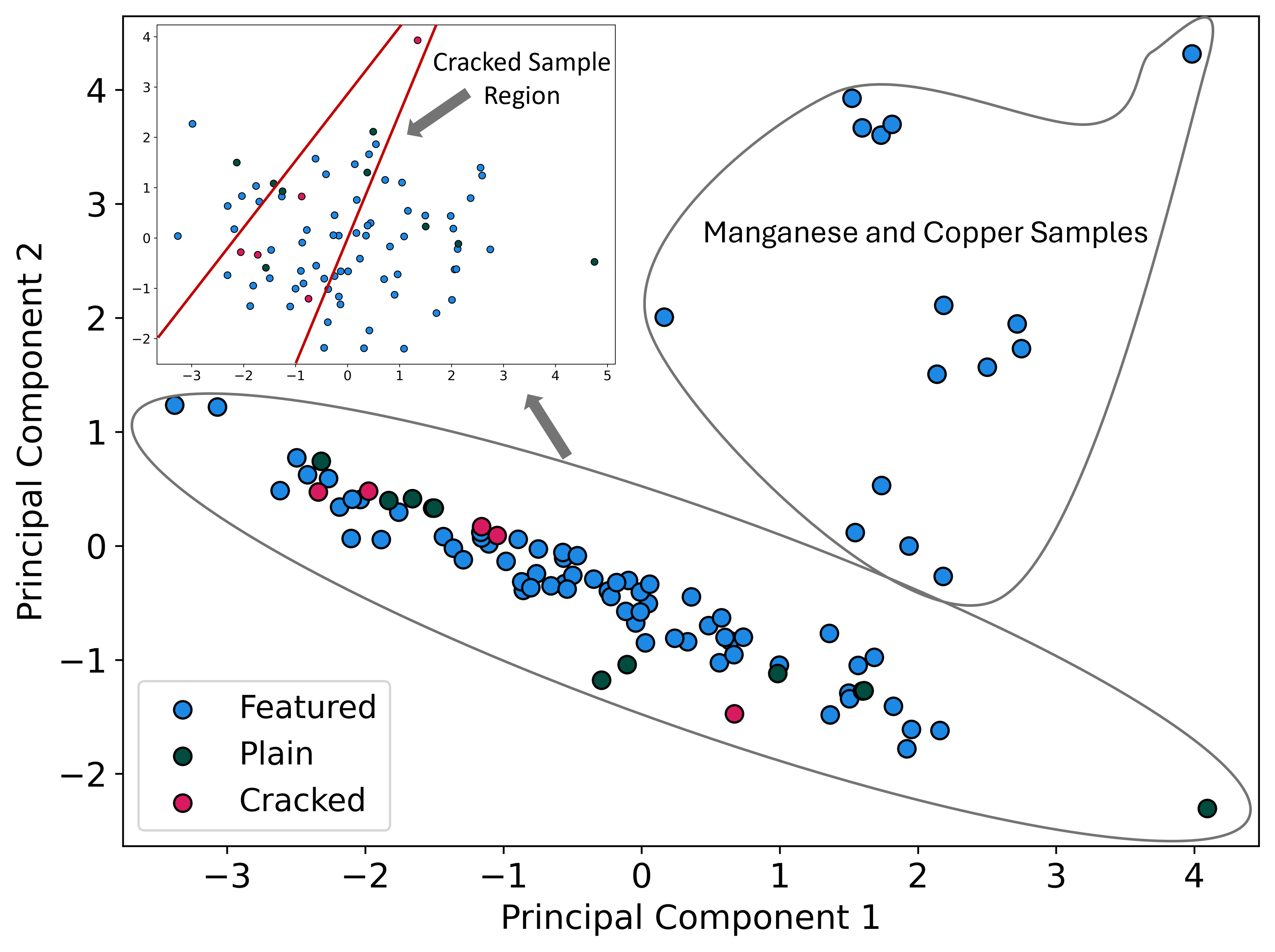}
        }%
        \hspace{3em}
        \subfloat[ANOVA analyses for Hardness in NI responses.]{%
         \includegraphics[width=0.45\textwidth, trim=0 0 0 50, clip]{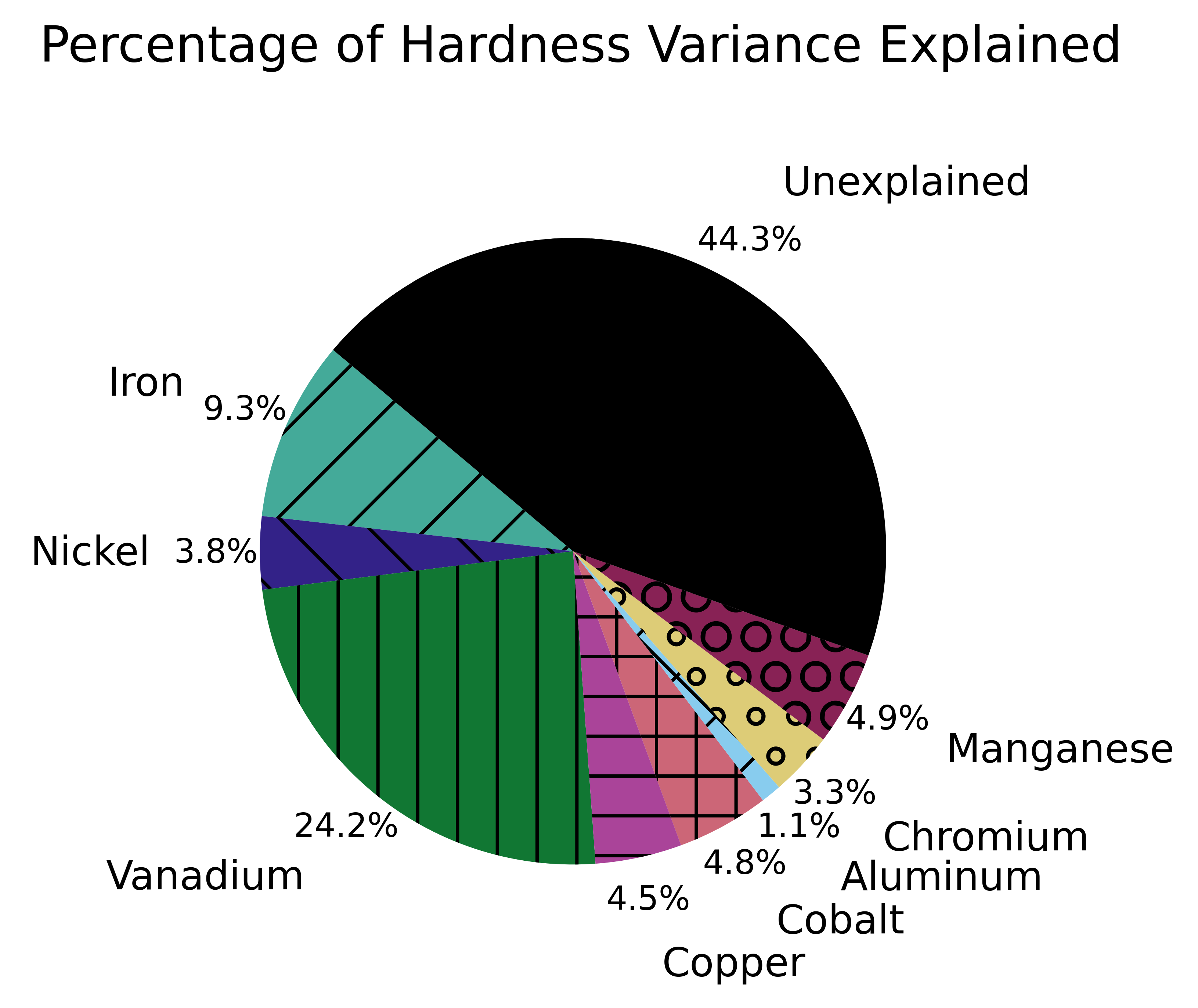}}        
    }
    
    \caption{Compositional sensitivity impact on nanoindentation response. (a) Various characteristic nanoindentation samples, showing three types of observed outcomes during testing. Sample AAB16 shows a pristine surface with no obvious deformation outside of the contact area, AAB10 shows pile–up and slip traces on the surface, and AAD16 exhibits brittle fracture. (b) SHAPley feature importance for composition effect on Nanoindentation cracking ML predictor results(Bi–class), ordered by magnitude. High Vanadium and low Nickel content in samples are seen to be the main causes of brittle indentation responses. (c) PCA reveals clustering of brittle samples in a local region, and samples including Manganese and Copper are far outside the initial domain. (d) ANOVA analyses reveal Vanadium and Iron are key contributors in influencing Hardness, in addition to a large unexplained variance present in hardness NI samples.}
    \label{fig:NI_data_analysis}
\end{figure*}

To better understand the underlying chemistry–property, and property-property relationships, an exploratory data analysis was conducted to assess correlations between elemental compositions and mechanical properties within each processing type. The scatter plot and correlation matrix in Figure~\ref{fig:Correlation Matrix} highlights key trends across the dataset, revealing both linear relationships and distributional characteristics. This visualization enables a detailed examination of feature interactions, variation in property distributions, and composition–property and property–property dependencies that inform subsequent modeling efforts.
Across both processing types, the highest experimental correlation is seen between Tension Elongation (\%) and UTS/YS (0.816), suggesting that alloys with a high UTS/YS will typically exhibit larger TE.
YS has a strong positive correlation with Hardness in hot forged samples (0.716) but also a strong negative correlation with UTS/YS (-0.826) in cold rolled materials, implying that there exists a trade-off between hard alloys with high yield strength and those with higher UTS/YS ratios.
Within hot forged samples, Vanadium is a key element as it has the highest chemistry–property correlations with Hardness (0.704), indicating materials with higher Vanadium content are likely to have higher hardness. 

\begin{figure*}[!ht]
    \centering    
    \includegraphics[width=.9\textwidth]{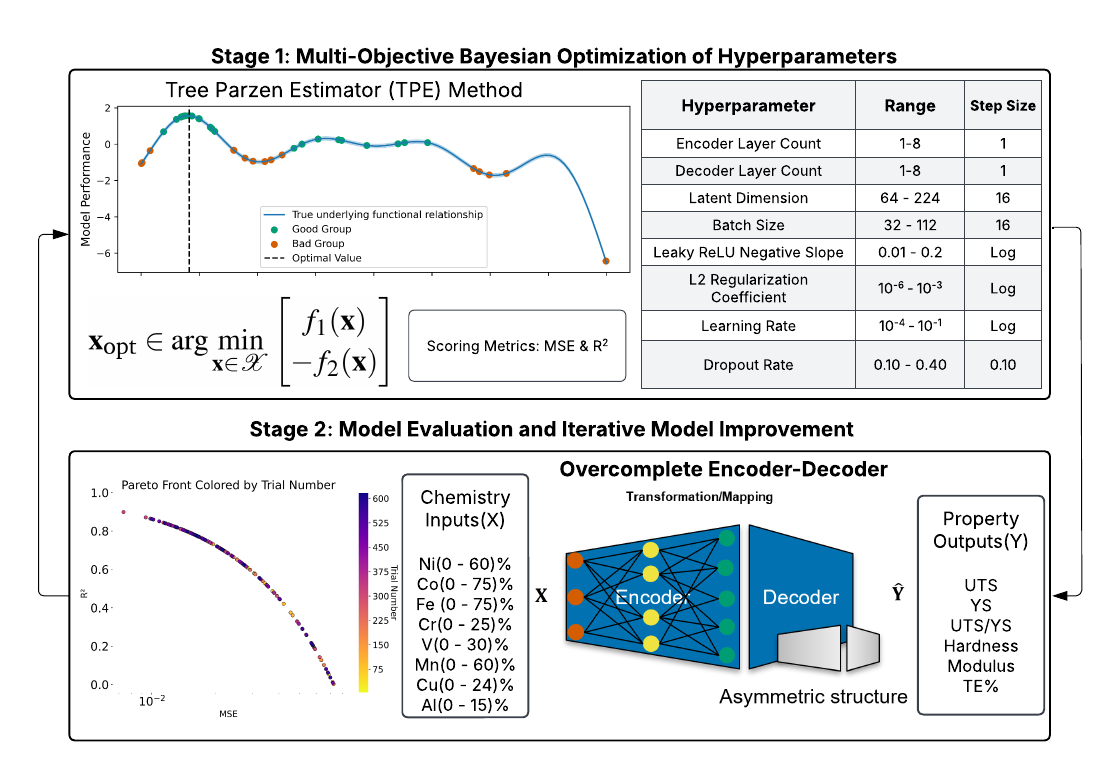}
    \caption{\textbf{Asymmetric overcomplete encoder–decoder model and multi–objective Bayesian hyperparameter optimization for learning HEA chemistry–property relationships}. Tree Parzen Estimator(TPE) minimization method applied over specified hyperparameter ranges for selected scoring metrics(top), followed by Encoder–Decoder property prediction and model evaluation(bottom), leading to iterative model improvement.`Log' step sizes are the use of a logarithmic algorithm sampling within the designated range.}    
    \label{fig:Encoder-Decoder-Model}
\end{figure*}

Building upon the insights gained from the exploratory correlation analysis, particular attention was directed toward understanding outlier mechanical behavior, specifically the occurrence of brittle deformation observed during nanoindentation testing. During the first year of BIRDSHOT alloy discovery campaign, a subset of samples displayed notable characteristics such as brittle cracking (5/96) or featureless indents (9/96), as shown in Figure~\ref{fig:NI_data_analysis}(a). These behaviors stand in contrast to the majority of samples that exhibited pile–up or twinning (82/96), indicative of more ductile responses. Since alloys exhibiting brittle behavior are generally unsuitable for high–performance structural applications, identifying compositional drivers of such responses is critical for guiding the design space in subsequent high–throughput experiments. A SHAP–based feature importance analysis shown in Figure ~\ref{fig:NI_data_analysis}(b) revealed that Vanadium and Nickel were the dominant contributors to a random forest classifier's predictions of brittleness. While prior works have applied SHAP strictly to compositional data for class prediction, this approach does not address inherent constraints of the simplex, where component features are strictly positive and sum to one, resulting in inherent collinearity \cite{noe2024explaining}. To address this constraint, we first apply the established isometric log ratio(ILR) transformation, mapping the data to an unconstrained Euclidean space \cite{egozcue2003isometric}.  After computing SHAP values in ILR space, we back-project them to the composition space using the sample-wise Jacobian of the ILR transformation. This local linear approximation maps SHAP attributions from the orthonormal ILR tangent space back onto the constrained simplex, enabling interpretation in terms of elemental compositions. This plot reveals that higher Vanadium content and low Nickel promote brittle class prediction. Notably, while vanadium was previously associated with higher hardness, its role here highlights a critical trade–off between strength–enhancing and embrittling effects, agreeing with past findings \cite{yin2020vanadium}. 

To further interpret the compositional trends underlying brittleness, a principal component analysis (PCA) was conducted, shown in Figure~\ref{fig:NI_data_analysis}(c). The first four components captured approximately 71\% of the total variance, indicating meaningful dimensionality reduction. Brittle samples clustered within a distinct region of the PCA space, characterized by low Nickel ($\leq35\%$) and high Vanadium ($\geq 20\%$ in 4 out of 5 cases). This pattern reinforces the SHAP–based findings and highlights a compositional regime associated with brittle behavior. Furthermore, all five brittle samples were found to deviate from the targeted single–phase FCC design, instead forming sigma phases, promoting multi-phase FCC(3/5) and D0$_{22}$ (1), or BCC structure(1), which likely contributed to their poor mechanical behavior. All sigma phase samples also exhibited brittle tensile testing results, reinforcing that brittle behavior in nanoindentation is predictive of poor tensile ductility and overall mechanical reliability. The original alloy designs were based TCHEA6 database predictions (using Thermocalc 2022.2), which indicated full FCC phase stability above 700\textdegree C for all candidate alloys under consideration. However, XRD analysis after synthesis revealed the formation of secondary phases not captured by TCHEA6. The updated TCHEA7 database now successfully predicts these secondary phases across all five alloys, providing a more accurate basis for future screening of brittle candidates and enhancing the reliability of single-phase FCC alloy design.

\begin{figure*}[!ht]
    \centering
    \tiny
    \begin{overpic}[width=0.31\textwidth]{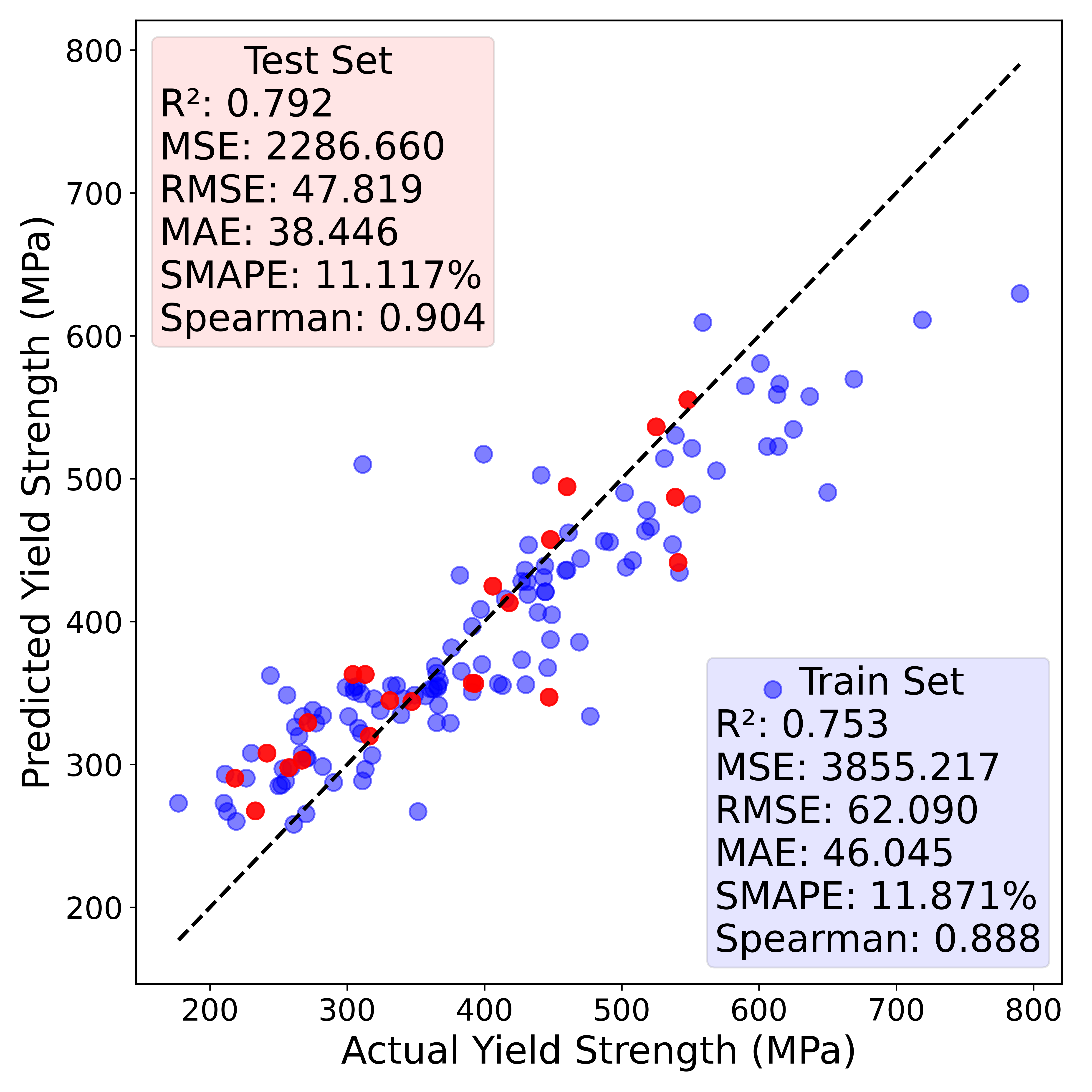}%
    \put(4,4){\large (a)}%
    \end{overpic}%
    \begin{overpic}[width=0.31\textwidth]{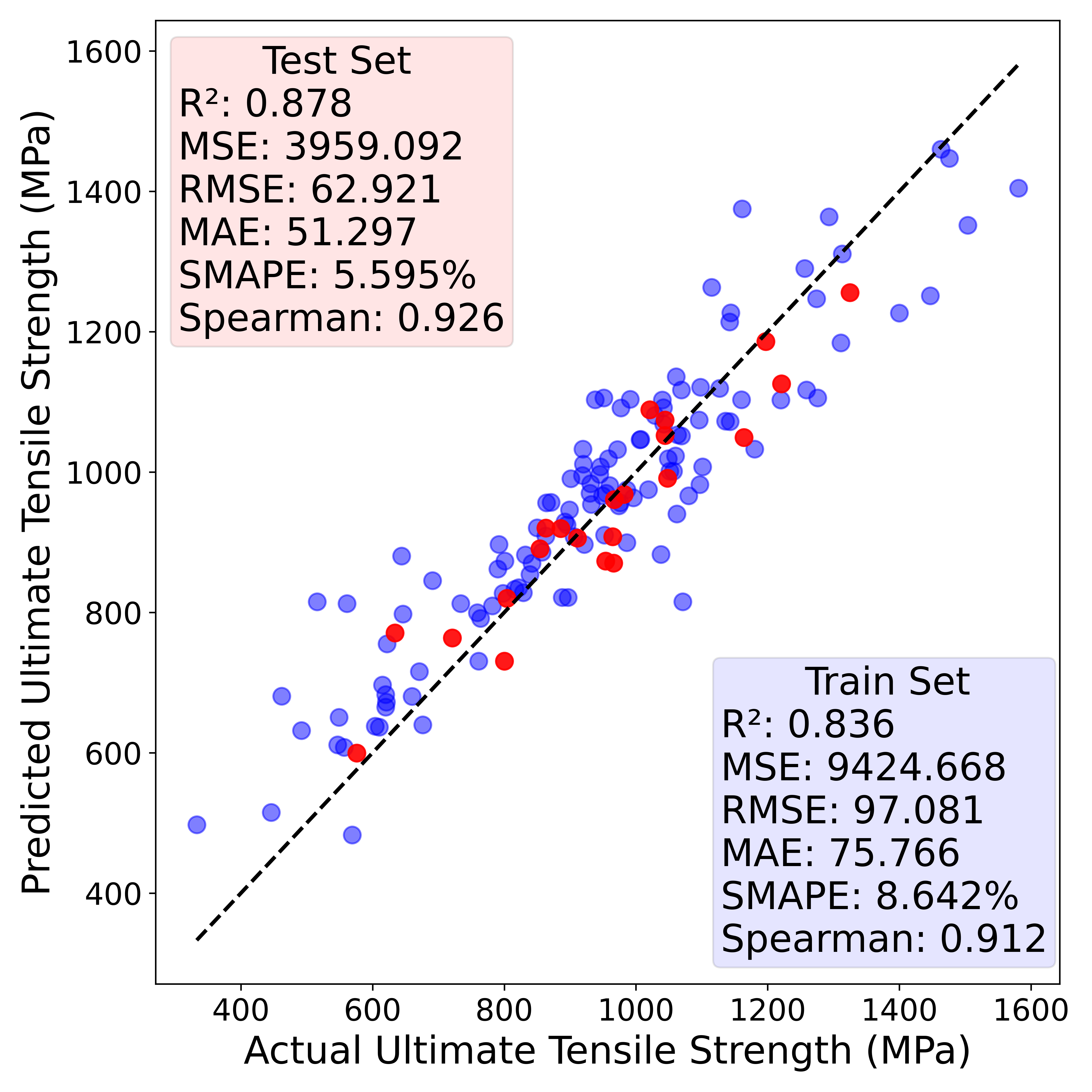}%
    \put(4,4){\large (b)}%
    \end{overpic}%
    \begin{overpic}[width=0.31\textwidth]{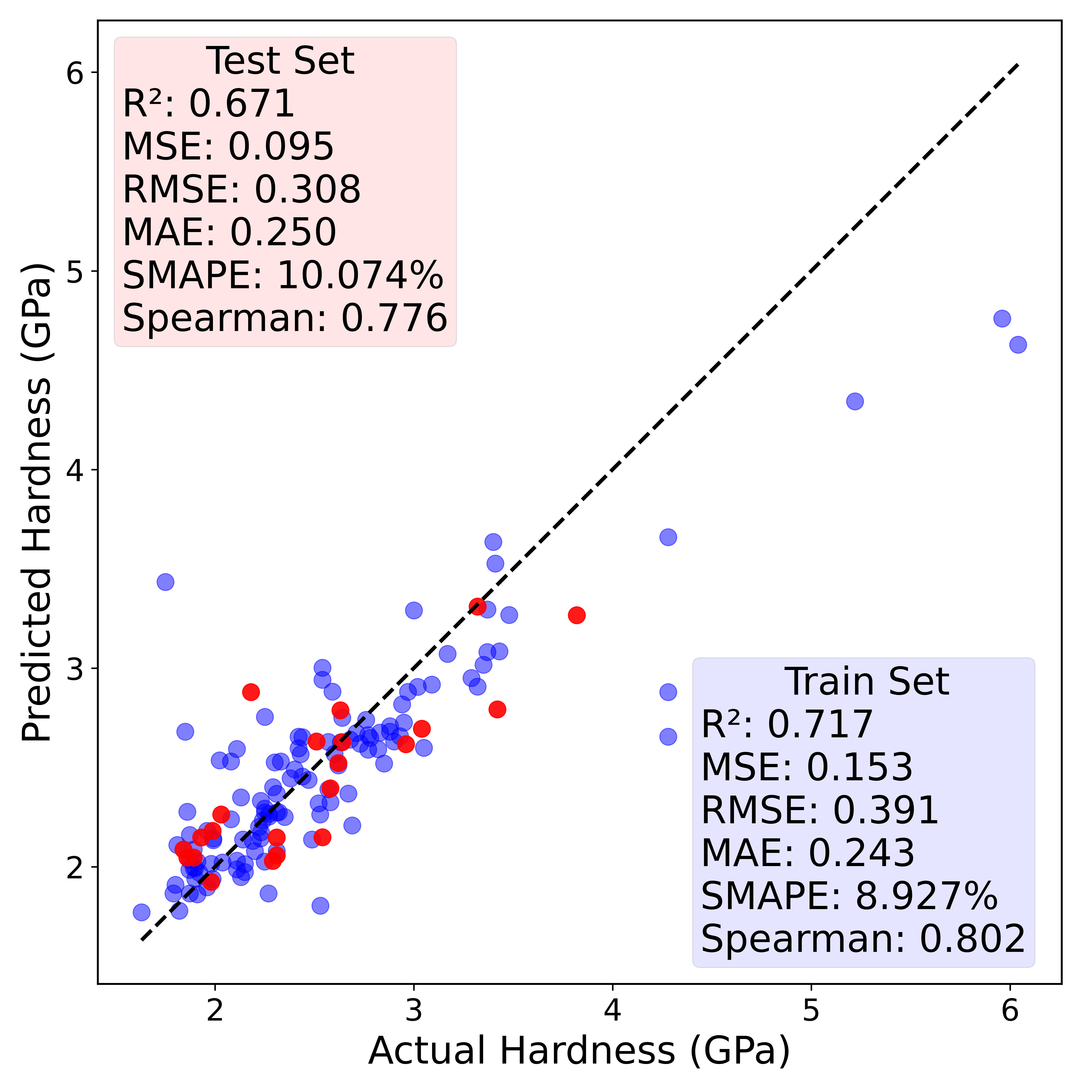}%
    \put(4,4){\large (c)}%
    \end{overpic}%
    \\
    \begin{overpic}[width=0.31\textwidth]{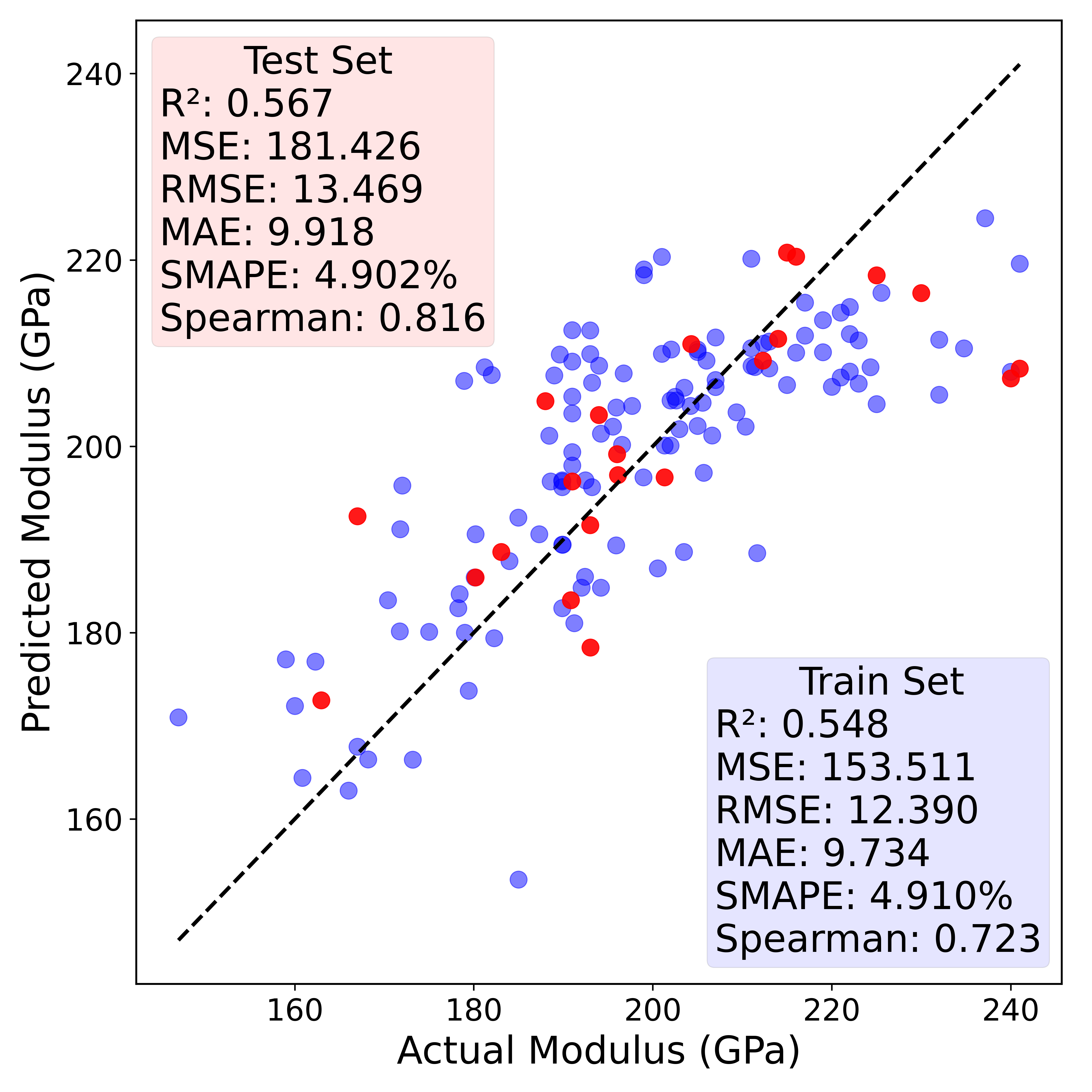}%
    \put(4,4){\large (d)}%
    \end{overpic}%
    \begin{overpic}[width=0.31\textwidth]{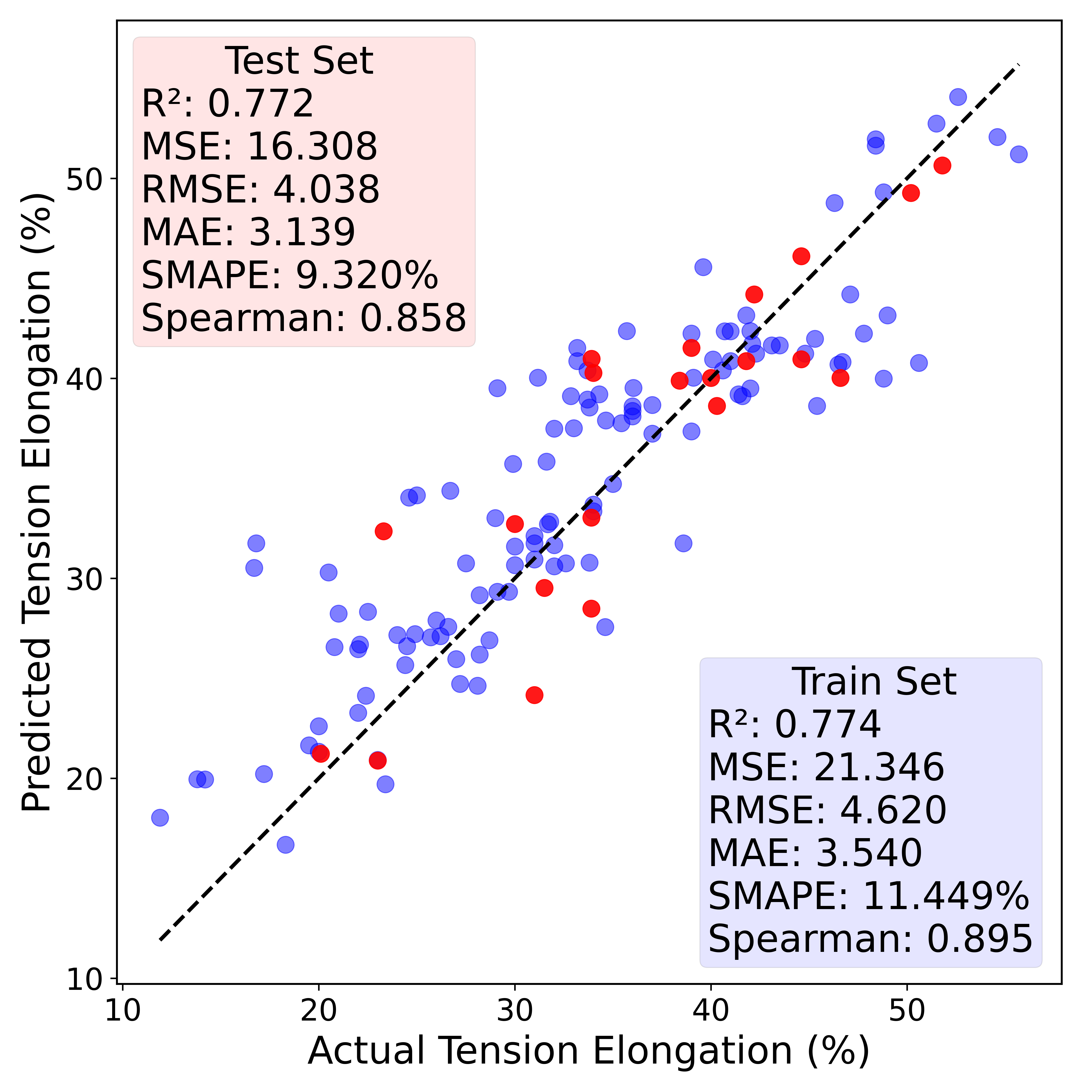}%
    \put(1,3){\large (e)}%
    \end{overpic}%
    \begin{overpic}[width=0.31\textwidth]{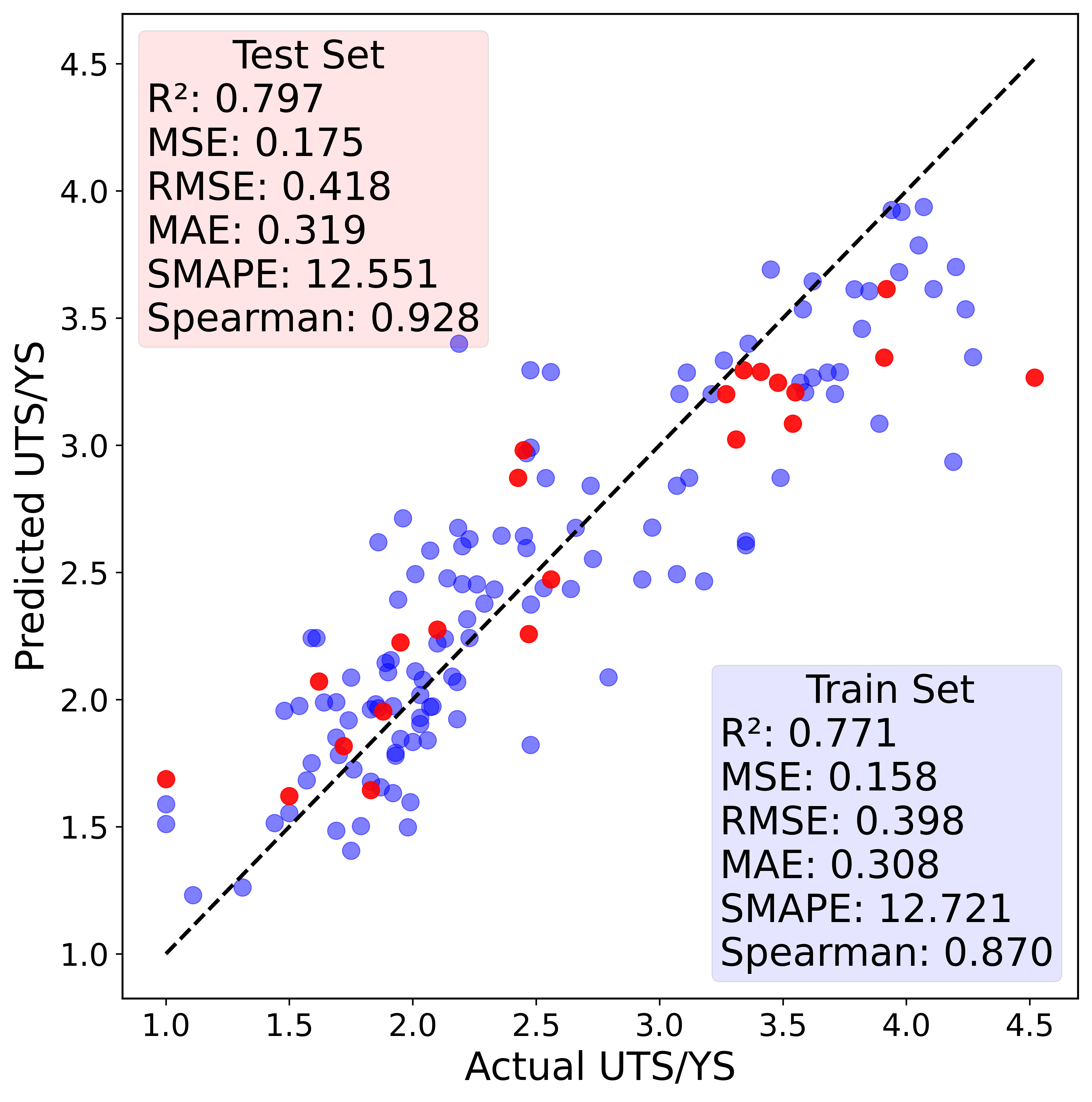}%
    \put(4,4){\large (f)}%
    \end{overpic}%
    \\

    \begin{overpic}[width=0.6\textwidth]{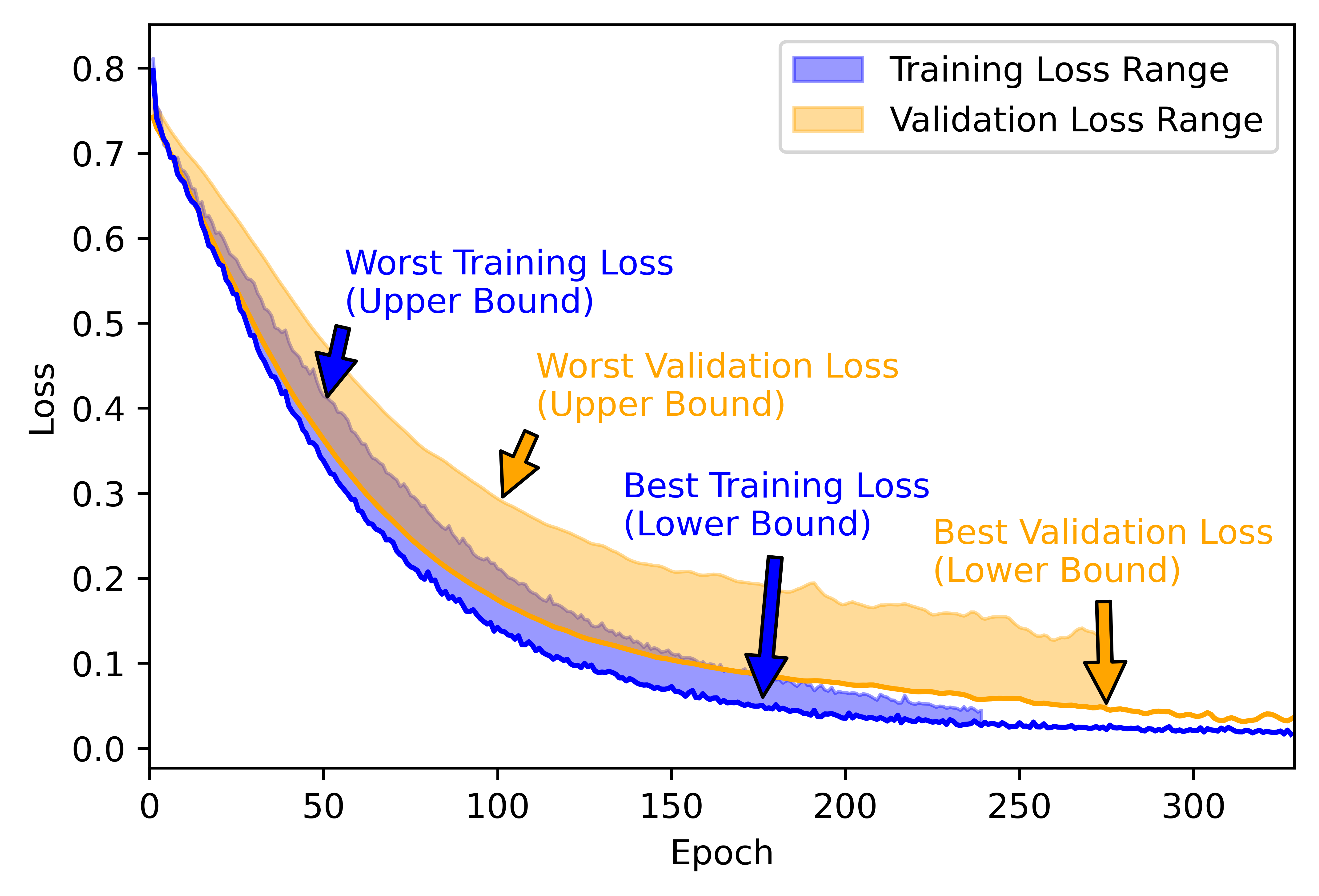}%
    \put(4,4){\large (g)}%
    \end{overpic}%

    \caption{\textbf{Parity plots showing the best predictive capability of the six Encoder–Decoder models for prediction of respective target outputs across 100 different test–train splits, when considering all scoring metrics for both the training and test sets.} We observe that in nearly all properties, there is slightly better performance in the testing set than in training, attributed to data sparsity and inherent model stochasticity. An Epoch–Loss training plot of the Individual UTS/YS predictor model reveals variation in model performance across these test–train splits.}
    \label{fig:HP-parity-plots}
\end{figure*}

To enhance the understanding of compositional influence in nanoindentation responses, a type II ANOVA analysis was conducted for Hardness, shown in Figure~\ref{fig:NI_data_analysis}(d). The four brittle sigma phase samples possessed the four highest hardness values in the dataset. Vanadium and Iron are seen as the key contributors in influencing measured hardness in NI, reinforcing their critical role in this system. Large unexplained variances in both properties likely stem from the local nature of nanoindentation experimental factors, including the relative size of the indenter and the affected crystal, variations in grain orientation, and the affected plane. Additionally, experimental noise in nanoindentation measurements can further reduce model accuracy~\cite{castillo2019bayesian}. A separate study on the other macroscopic properties, such as UTS, revealed less unexplained variance. 

\begin{table*}[ht]
    \centering
    \scriptsize
    \begin{tabular}{|c |c c c c c c c c|}
        \hline
\textbf{Property} & 
\textbf{\shortstack{Encoder\\Layers}} & 
\textbf{\shortstack{Decoder\\Layers}} &  
\textbf{\shortstack{Latent\\Dimension}} & 
\textbf{\shortstack{Dropout (\%)}} & 
\textbf{\shortstack{Leaky ReLU\\Negative Slope}} & 
\textbf{\shortstack{L2 Regularization\\Coefficient}} & 
\textbf{\shortstack{Batch\\Size}} &
\textbf{\shortstack{Learning\\Rate}} \\
\hline

        Yield Strength & 4 & 4 & 96 & 10 & 0.091 & $1.11 \times 10^{-4}$ & 68 & $1.33\times10^{-3}$\\
        UTS & 3 & 2 & 192 & 10 & 0.136 & $2.07\times10^{-4}$ & 44 & $8.47\times10^{-4}$\\
        TE \% & 1 & 1 & 208 & 40 & 0.014 & $1.35\times10^{-3}$ & 38 & $1.97\times10^{-4}$\\
        Hardness & 2 & 2 & 200 & 30 & 0.019 & $6.64\times10^{-4}$ & 64 & $1.09\times10^{-3}$\\
        Modulus & 2 & 1 & 98 & 30 & 0.014 & $8.73\times10^{-4}$ & 78 & $1.53\times10^{-4}$\\
        UTS/YS Ratio & 1 & 1 & 176 & 20 & 0.111 & $9.87\times10^{-3}$ & 54 & $1.66\times10^{-4}$\\
        \hline
    \end{tabular}
    \caption{Observed optimal Encoder–Decoder hyperparameters for modeling individual properties after 600 iterations of BO and hyperparameter search bound refining.}
    \label{tab:best_hparams}
\end{table*}

To enhance predictive accuracy in sparsely sampled alloy spaces, the BIRDSHOT BO framework incorporates prior estimates of material properties. In BO, a prior model provides an initial belief about the objective function, typically via a Gaussian Process (GP) or pretrained ML model, guiding early exploration before new data is observed. While standard BO assumes a zero–mean GP, incorporating a model such as an encoder–decoder neural network enables a non–zero mean function, allowing the GP to learn residuals relative to the prior, improving sample efficiency and accelerating convergence, critical advantages in data–scarce materials discovery settings.

Accurate prediction of HEA properties in the context of BO is hindered by the limited size of experimental datasets and the inherently complex, high–dimensional nature of compositional design spaces. Tree–based models such as CatBoost and XGBoost are commonly employed in low–data regimes (e.g., <200 samples) for their robustness and ease of use~\cite{ansari2022comprehensive, barrionuevo2021comparative, khosravi2022performance}, but they may struggle to fully capture the intricate, non–linear interactions present in HEAs without extensive feature engineering. Neural networks, by contrast, offer greater representational capacity for capturing complex, non–linear relationships, though this often comes at the expense of model interpretability~\cite{ilyas2019adversarial, zhong2022explainable, montavon2018methods}. However, their application is typically constrained by the need for large training datasets, often unavailable in early–stage materials discovery. In this study, we address this limitation by employing Bayesian hyperparameter optimization to tune neural architectures effectively under data scarcity. This approach enables improved regularization, efficient architecture selection, and enhanced generalization, allowing neural networks to outperform conventional regressors in predicting key materials properties of HEAs, even with limited training data.


We implement a sparsely connected, asymmetric overcomplete encoder-decoder neural network to model chemistry–property relationships in HEAs. This choice is motivated by the high dimensionality and compositional complexity of the dataset, which challenge conventional regression models. In the encoder-decoder architecture, input compositions are projected into a higher–dimensional latent space, allowing the network to extract more abstract and informative representations. Overcompleteness, where the latent dimensionality exceeds that of the input, enables the model to capture subtle, non–linear correlations that are often missed by simpler models. To prevent the network from learning trivial identity mappings, we impose sparsity on the connections. This acts as a structural constraint, encouraging generalization and promoting the emergence of disentangled, physically meaningful features. The architecture originally proposed in~\cite{attari2024decoding} has proven effective for modeling complex materials behavior in large materials datasets; in this work, we adapt it to the data–scarce setting typical of early–stage alloy discovery. Within the BO framework (Figure~\ref{fig:Encoder-Decoder-Model}), this model serves as the backbone for evaluating performance across a range of architectural and regularization configurations.

Hyperparameter tuning is essential for optimizing neural network performance, particularly given the architectural complexity and sensitivity of such models~\cite{weerts2020importance}. Manual tuning and grid search become impractical in high–dimensional, interdependent parameter spaces due to computational cost and inefficiency~\cite{wang2023scientific}. To address this, we employ BO using the Tree–structured Parzen Estimator (TPE)~\cite{eggensperger2013towards, akiba2019optuna}, which models the distributions of high– and low–performing configurations to balance exploration and exploitation. TPE is particularly well–suited for neural architectures with conditional and mixed–type hyperparameters. We apply it to tune key architectural variables of our encoder–decoder model, including the number of layers, learning rate, and regularization strength. Optimization is guided by two performance metrics, calculated from the test set predictions: mean squared error (MSE), which is minimized, and coefficient of determination (R²), which is maximized. Minimizing MSE discourages extreme mispredictions, while simultaneously maximizing R² promotes capturing the underlying structure in the data. The implemented Pareto front approach considers both metrics equally, searching for configurations that are non–dominated in both objectives. All sampled hyperparameter configurations and their evaluation scores are retained for reproducibility and post–hoc analysis.

\begin{table*}[htbp]
    \centering
    \scriptsize
    \begin{tabular}{|c| c c| c c| c c c c|}
        \hline
        \textbf{Property} & \textbf{Skewness} & \textbf{Kurtosis} & \textbf{\shortstack{E–D Train\\Mean (Best)}} & \textbf{\shortstack{Mutli–task Train\\Mean (Best)}} & \textbf{\shortstack{E–D Test\\Mean (Best)}} & \textbf{\shortstack{Multi–task Test\\Mean (Best)}} & \textbf{\shortstack{XGBoost Test\\Mean (Best)}} & \textbf{\shortstack{Catboost Test\\Mean (Best)}}\\
        \hline
        
        UTS & 0.06 & 0.13 & 0.682 \textbf{\underline{(0.910)}} & 0.768 (0.849) & 0.398 \textbf{\underline{(0.870)}} & \textbf{\underline{0.443}} (0.745) & 0.348 (0.642) & 0.426 (0.781)\\

        YS & 0.58 & -0.17 & 0.346 (0.756) & \textbf{\underline{0.792}} \textbf{\underline{(0.868)}} & 0.336 \textbf{\underline{(0.872)}} & \textbf{\underline{0.507}} (0.807) & 0.503 (0.816) & 0.477 (0.827)\\

        UTS/YS Ratio & 0.50 & -0.85 & 0.786 (0.854) & \textbf{\underline{0.810}} \textbf{\underline{(0.888)}} & 0.511 (0.841) & 0.525 \textbf{\underline{(0.851)}} & \textbf{\underline{0.557}} (0.799) & 0.517 (0.802)\\
        
        TE \% & -0.61 & 0.61 & 0.386 (0.817) & \textbf{\underline{0.779}} \textbf{\underline{(0.854)}} & 0.348 \textbf{\underline{(0.855)}} & \textbf{\underline{0.479}} (0.827) & 0.423 (0.726) & 0.458 (0.768)\\
        
        Hardness & 2.3 & 7.62 & 0.592 (0.799) & \textbf{\underline{0.710}} \textbf{\underline{(0.819)}} & \textbf{\underline{0.349}} (0.756) & 0.317 (0.780) & 0.311 \textbf{\underline{(0.799)}} & 0.158 (0.777)\\
        
        Modulus & 0.01 & -0.12 & 0.405 (0.519) & \textbf{\underline{0.653}} \textbf{\underline{(0.743)}} & 0.197 (0.567) & 0.178 \textbf{\underline{(0.609)}} & \textbf{\underline{0.279}} (0.578) & 0.264 (0.517)\\
        
        \hline
    \end{tabular}
    \caption{Property distribution quantification and comparison of mean (first value) and best  R² (enclosed second value) prediction metric scores for each model across all properties. The best hyperparameters observed in Table 1 are implemented, and these models are used across 100 different train–test splits, in addition to XGBoost and Catboost. 
    A grid search hyperparameter tuning was performed for both XGBoost and Catboost to compare the best results, and both conventional regressors predicted properties individually. 
    E–D shows Encoder–Decoder prediction for single property prediction, while the multi–task model, which predicts all properties simultaneously, shows several improved results compared to individual predictor results. The average multi–task model performance is improved for nearly all metrics, and the best results remain similar, showing that the model learning the correlations in property–property data improves the mean prediction performance.}
    \label{tab:Regressor_comparison_table}
\end{table*}

This framework was applied to a sparsely connected, overcomplete encoder–decoder model to predict six mechanical properties individually. Although multi–task prediction is theoretically possible, we chose single–property models to better capture distinct chemistry–property relationships. Each model was trained using the Adam optimizer and the MSE loss function, which outperformed alternatives in initial benchmarking. An 85/15 train–test split was used for accurate data-sparse modeling, and model performance was evaluated using both MSE and R² metrics. The total optimization included 600 trials, with hyperparameter bounds adjusted after 150, 300, and 450 trials based on Optuna's importance scores and contour visualizations. Beyond 600 trials, performance gains plateaued.

The final optimal configurations, seen in Table~\ref{tab:best_hparams}, revealed that effective models typically had no more than four layers per sub–network and employed various dropout rates. Figure~\ref{fig:HP-parity-plots} presents parity plots comparing predicted versus actual values for six mechanical properties of HEAs using the encoder–decoder model. Each subplot corresponds to one target property: (a) Yield Strength, (b) Ultimate Tensile Strength, (c) Hardness, (d) Modulus, (e) Tension Elongation, and (f) UTS/YS ratio. Red markers denote the test set, while blue markers indicate training data. Six evaluation metrics for both the training and testing sets are included. MSE = Mean Squared Error, RMSE = Root Mean Squared Error, MAE = Mean Absolute Error, SMAPE = Symmetric Mean Absolute Percentage Error, and Spearman reflects the linear correlation coefficient. Considering the small size of dataset, across all properties, the model achieves strong predictive performance, with test set R² values ranging from 0.567 (Modulus) to 0.878 (UTS), and Spearman correlation coefficients above 0.81 in each case. The model generalizes well across all properties despite being trained on sparse data, as evidenced by the close clustering of test points around the ideal diagonal line. These results demonstrate the model's capability to effectively learn complex composition–property relationships. The model exhibits strong predictive capability in measuring macroscopic properties, which are better predicted by chemistry. The most notable prediction errors occur in modulus and extreme hardness values, where the model struggles to predict extreme values. The increased scatter and lower R² of these properties likely stem from the previously mentioned local nature of nanoindentation measurements. Minor deviations are also observed in UTS/YS and TE predictions at the extremes, while YS and UTS predictions show minimal error and strong alignment with the parity line, reflecting higher model confidence in these properties. Note that the BIRDSHOT center data used for model training was current as of December 2024.

Results from the best models (Figure~\ref{fig:HP-parity-plots}, Table~\ref{tab:Regressor_comparison_table}) show that a Bayesian–optimized encoder–decoder model can outperform conventional regressors in predicting HEA properties using sparse data. To assess robustness, the UTS/YS model was retrained on 100 independent train–test splits. The resulting performance variability is shown in Figure~\ref{fig:HP-parity-plots}(g), confirming stability and generalization across data partitions.

To test the intuition that creating individual models is necessary for optimal prediction performance, a model showing high predictive ability, the UTS model, was implemented as a multi–task predictor to predict all 6 properties simultaneously. When applied in a multi-task setting, the model was able to learn the underlying property–property correlations in the data, improving the mean predictive performance in both training and test sets, as seen in Table ~\ref{tab:Regressor_comparison_table}. The model outperformed the individual models' mean predictive capabilities in the test set for all tension properties, while also decreasing prediction standard deviation, even in a small dataset. The multi–task predictor exhibited increased poor outlier performances in comparison to the single predictor models for 4/6 properties, showcasing the hyperparameter sensitivity of the multi–task predictor. While the multi–task model does achieve higher predictive performance in both NI properties, it is reduced in 3/4 tension properties, establishing that there are tradeoffs when considering this approach. When considering both single and multi–task prediction results, conventional regressors are outperformed in 9/12 comparison scores.

In conclusion, sensitivity analyses revealed key chemistry–property relationships in nanoindentation samples, and BO was implemented to tune data–driven overcomplete Encoder–Decoder models, drastically improving their utility in HEA property prediction, even with sparse data. Various analyses reveal significant correlations between composition and mechanical properties, including Vanadium's pronounced influence in nanoindentation responses. The demonstrated Bayesian hyperparameter tuning framework reveals optimal Encoder–Decoder configurations, capable of outperforming conventional regressors in the property prediction of HEAs. A tuned model, employed as a multi–task predictor, leveraged inter-property correlations to increase both mean predictive performance and reduce prediction variance, enhancing its overall predictive power for several properties. However, without the integration of domain knowledge and physical principles, purely data–driven models risk producing unreliable or non–generalizable predictions. A hybrid approach that combines ML with materials science insight is essential for truly impactful HEA design.

\section*{Conflicts of interest}
The authors declare that they have no known competing financial interests or personal relationships that could have appeared to influence the work reported in this paper.

\section*{Code and Data availability}
The code and data associated with this work are both available at the following repository: https://doi.org/10.24433/CO.9160143.v1

\section*{Acknowledgments}
The authors acknowledge the support from DEVCOM–ARL under Contract No. W911NF-22-2-0106 (\emph{BIRDSHOT} Center funded through the High–throughput Materials Discovery for Extreme Conditions (HTMDEC)). 


\sloppy
\bibliography{mybibfile}

\begin{thebibliography}{10}

\bibitem{cantor2004microstructural}
Brain Cantor, ITH Chang, Peter Knight, and AJB Vincent.
\newblock Microstructural development in equiatomic multicomponent alloys.
\newblock {\em Materials Science and Engineering: A}, 375:213--218, 2004.

\bibitem{yeh2004nanostructured}
J-W Yeh, S-K Chen, S-J Lin, J-Y Gan, T-S Chin, T-T Shun, C-H Tsau, and S-Y Chang.
\newblock Nanostructured high-entropy alloys with multiple principal elements: novel alloy design concepts and outcomes.
\newblock {\em Advanced engineering materials}, 6(5):299--303, 2004.

\bibitem{mulukutla2024illustrating}
Mrinalini Mulukutla, A~Nicole Person, Sven Voigt, Lindsey Kuettner, Branden Kappes, Danial Khatamsaz, Robert Robinson, Daniel~Salas Mula, Wenle Xu, Daniel Lewis, et~al.
\newblock Illustrating an effective workflow for accelerated materials discovery.
\newblock {\em Integrating Materials and Manufacturing Innovation}, pages 1--21, 2024.

\bibitem{borg2020expanded}
Christopher~KH Borg, Carolina Frey, Jasper Moh, Tresa~M Pollock, St{\'e}phane Gorsse, Daniel~B Miracle, Oleg~N Senkov, Bryce Meredig, and James~E Saal.
\newblock Expanded dataset of mechanical properties and observed phases of multi-principal element alloys.
\newblock {\em Scientific Data}, 7(1):430, 2020.

\bibitem{allen2024performance}
Marshall~D Allen, Vahid Attari, Brent Vela, James Hanagan, Richard Malak, and Raymundo Arr{\'o}yave.
\newblock Performance-driven computational design of multi-terminal compositionally graded alloy structures using graphs.
\newblock {\em arXiv preprint arXiv:2412.03674}, 2024.

\bibitem{khatamsaz2022multi}
Danial Khatamsaz, Brent Vela, Prashant Singh, Duane~D Johnson, Douglas Allaire, and Raymundo Arr{\'o}yave.
\newblock Multi-objective materials bayesian optimization with active learning of design constraints: Design of ductile refractory multi-principal-element alloys.
\newblock {\em Acta Materialia}, 236:118133, 2022.

\bibitem{khatamsaz2023bayesian}
Danial Khatamsaz, Brent Vela, Prashant Singh, Duane~D Johnson, Douglas Allaire, and Raymundo Arr{\'o}yave.
\newblock Bayesian optimization with active learning of design constraints using an entropy-based approach.
\newblock {\em npj Computational Materials}, 9(1):49, 2023.

\bibitem{hastings2024interoperable}
Trevor Hastings, Mrinalini Mulukutla, Danial Khatamsaz, Daniel Salas, Wenle Xu, Daniel Lewis, Nicole Person, Matthew Skokan, Braden Miller, James Paramore, et~al.
\newblock An interoperable multi objective batch bayesian optimization framework for high throughput materials discovery.
\newblock {\em arXiv preprint arXiv:2405.08900}, 2024.

\bibitem{demvsar2004orange}
Janez Dem{\v{s}}ar, Bla{\v{z}} Zupan, Gregor Leban, and Tomaz Curk.
\newblock Orange: From experimental machine learning to interactive data mining.
\newblock In {\em Knowledge Discovery in Databases: PKDD 2004: 8th European Conference on Principles and Practice of Knowledge Discovery in Databases, Pisa, Italy, September 20-24, 2004. Proceedings 8}, pages 537--539. Springer, 2004.

\bibitem{noe2024explaining}
Paul-Gauthier No{\'e}, Miquel Perell{\'o}-Nieto, Jean-Fran{\c{c}}ois Bonastre, and Peter Flach.
\newblock Explaining a probabilistic prediction on the simplex with shapley compositions.
\newblock In {\em ECAI 2024}, pages 1124--1131. IOS Press, 2024.

\bibitem{egozcue2003isometric}
Juan~Jos{\'e} Egozcue, Vera Pawlowsky-Glahn, Gl{\`o}ria Mateu-Figueras, and Carles Barcelo-Vidal.
\newblock Isometric logratio transformations for compositional data analysis.
\newblock {\em Mathematical geology}, 35(3):279--300, 2003.

\bibitem{yin2020vanadium}
Binglun Yin, Francesco Maresca, and William~A Curtin.
\newblock Vanadium is an optimal element for strengthening in both fcc and bcc high-entropy alloys.
\newblock {\em Acta Materialia}, 188:486--491, 2020.

\bibitem{castillo2019bayesian}
Andrew Castillo and Surya~R Kalidindi.
\newblock A bayesian framework for the estimation of the single crystal elastic parameters from spherical indentation stress-strain measurements.
\newblock {\em Frontiers in Materials}, 6:136, 2019.

\bibitem{ansari2022comprehensive}
Sam Ansari and Ali~Bou Nassif.
\newblock A comprehensive study of regression analysis and the existing techniques.
\newblock In {\em 2022 Advances in Science and Engineering Technology International Conferences (ASET)}, pages 1--10. IEEE, 2022.

\bibitem{barrionuevo2021comparative}
German~Omar Barrionuevo, Sergio Rios, Stewart~W Williams, and Jorge~Andres Ramos-Grez.
\newblock Comparative evaluation of machine learning regressors for the layer geometry prediction in wire arc additive manufacturing.
\newblock In {\em 2021 IEEE 12th International Conference on Mechanical and Intelligent Manufacturing Technologies (ICMIMT)}, pages 186--190. IEEE, 2021.

\bibitem{khosravi2022performance}
Marzieh Khosravi, Sadman~Bin Arif, Ali Ghaseminejad, Hamed Tohidi, and Hanieh Shabanian.
\newblock Performance evaluation of machine learning regressors for estimating real estate house prices.
\newblock {\em Preprints}, page 2022090341, 2022.

\bibitem{ilyas2019adversarial}
Andrew Ilyas, Shibani Santurkar, Dimitris Tsipras, Logan Engstrom, Brandon Tran, and Aleksander Madry.
\newblock Adversarial examples are not bugs, they are features.
\newblock {\em Advances in neural information processing systems}, 32, 2019.

\bibitem{zhong2022explainable}
Xiaoting Zhong, Brian Gallagher, Shusen Liu, Bhavya Kailkhura, Anna Hiszpanski, and T~Yong-Jin Han.
\newblock Explainable machine learning in materials science.
\newblock {\em npj computational materials}, 8(1):204, 2022.

\bibitem{montavon2018methods}
Gr{\'e}goire Montavon, Wojciech Samek, and Klaus-Robert M{\"u}ller.
\newblock Methods for interpreting and understanding deep neural networks.
\newblock {\em Digital signal processing}, 73:1--15, 2018.

\bibitem{attari2024decoding}
Vahid Attari and Raymundo Arroyave.
\newblock Decoding non-linearity and complexity: deep tabular learning approaches for materials science.
\newblock {\em Digital Discovery}, 2025.

\bibitem{weerts2020importance}
Hilde~JP Weerts, Andreas~C Mueller, and Joaquin Vanschoren.
\newblock Importance of tuning hyperparameters of machine learning algorithms.
\newblock {\em arXiv preprint arXiv:2007.07588}, 2020.

\bibitem{wang2023scientific}
Hanchen Wang, Tianfan Fu, Yuanqi Du, Wenhao Gao, Kexin Huang, Ziming Liu, Payal Chandak, Shengchao Liu, Peter Van~Katwyk, Andreea Deac, et~al.
\newblock Scientific discovery in the age of artificial intelligence.
\newblock {\em Nature}, 620(7972):47--60, 2023.

\bibitem{eggensperger2013towards}
Katharina Eggensperger, Matthias Feurer, Frank Hutter, James Bergstra, Jasper Snoek, Holger Hoos, Kevin Leyton-Brown, et~al.
\newblock Towards an empirical foundation for assessing bayesian optimization of hyperparameters.
\newblock In {\em NIPS workshop on Bayesian Optimization in Theory and Practice}, volume~10, pages 1--5, 2013.

\bibitem{akiba2019optuna}
Takuya Akiba, Shotaro Sano, Toshihiko Yanase, Takeru Ohta, and Masanori Koyama.
\newblock Optuna: A next-generation hyperparameter optimization framework.
\newblock In {\em Proceedings of the 25th ACM SIGKDD international conference on knowledge discovery \& data mining}, pages 2623--2631, 2019.

\end{thebibliography}

\end{document}